# Rich magnetoelectric phase diagrams of multiferroic single-crystal α-NaFeO$_2$

Noriki Terada,[1,*] Yuta Ikedo,[2] Hirohiko Sato,[2,†] Dmitry D. Khalyavin,[3] Pascal Manuel,[3] Atsushi Miyake,[4] Akira Matsuo,[4] Masashi Tokunaga,[4] and Koichi Kindo[4]

[1]*National Institute for Materials Science, Sengen 1-2-1, Tsukuba, Ibaraki 305-0047, Japan*
[2]*Department of Physics, Chuo University, 1-13-27 Kasuga, Bunkyo-ku, Tokyo 112-8551, Japan*
[3]*ISIS Facility, STFC Rutherford Appleton Laboratory, Chilton, Didcot, Oxfordshire OX11 0QX, United Kingdom*
[4]*Institute for Solid State Physics, University of Tokyo, 5-1-5 Kashiwanoha, Kashiwa, Chiba 277-8581, Japan*

The magnetic and dielectric properties of the multiferroic triangular lattice magnet compound α-NaFeO$_2$ were studied by magnetization, specific heat, dielectric permittivity, and pyroelectric current measurements and by neutron diffraction experiments using single crystals grown by a hydrothermal synthesis method. This work produced magnetic field (in the monoclinic $ab$-plane, $B_{ab}$, and along the $c^*$-axis, $B_c$) versus temperature magnetic phase diagrams, including five and six magnetically ordered phases in $B_{ab}$ and along $B_c$, respectively. In zero magnetic field, two spin-density-wave orderings with different $\bm{k}$ vectors—$(0,q,\frac{1}{2})$ in phase I and $(q_a,q_b,q_c)$ in phase II—appeared at $T = 9.5$ and 8.25 K, respectively. Below $T = 5$ K, a commensurate order with $\bm{k} = (0.5,0,0.5)$ was stabilized as the ground state in phase III. Both $B_{ab} \geqslant 3$ T and $B_c \geqslant 5$ T were found to induce ferroelectric phases at the lowest temperature (2 K), with an electric polarization that was not confined to any highly symmetric directions in phases IV$_{ab}$ ($3.3 \leqslant B_{ab} \leqslant 8.5$ T), V$_{ab}$ ($8.5 \leqslant B_{ab} \leqslant 13.6$ T), IV$_c$ ($5.0 \leqslant B_c \leqslant 8.5$ T), and V$_c$ ($8.5 \leqslant B_c \leqslant 13.5$ T). In phase VI$_c$, within a narrow temperature region in $B_c$, the polarization was confined to the $ab$ plane. For each of the ferroelectric phases, the $\bm{k}$ vector was $(q_a,q_b,q_c)$, and noncollinear structures were identified, including a general spiral in IV$_{ab}$, an $ab$ cycloid in IV$_c$ and V$_c$, and a proper screw in VI$_c$, along with a triclinic 11' magnetic point group allowing polarization in the general direction. Comparing the polarization direction to the magnetic structures in the ferroelectric phases, we conclude that the extended inverse Dzyaloshinskii-Moriya mechanism expressed by the orthogonal components $\bm{p}_1 \propto \bm{r}_{ij} \times (\bm{S}_i \times \bm{S}_j)$ and $\bm{p}_2 \propto \bm{S}_i \times \bm{S}_j$ can explain the polarization directions. Based on calculations incorporating exchange interactions up to fourth-nearest-neighbor (NN) couplings, we infer that competition among antiferromagnetic second NN interactions in the triangular lattice plane, as well as weak interplane antiferromagnetic interactions, are responsible for the rich phase diagrams of α-NaFeO$_2$.

## I. INTRODUCTION

Over the past decade, the relationship between magnetic and ferroelectric orderings in multiferroic materials has been studied extensively so as to understand their novel magnetoelectric coupling [1–3]. In most spin-order driven multiferroics, so-called type-II multiferroics, an incommensurate noncollinear spin ordering stabilized as the result of spin frustration breaks spatial inversion symmetry. In some theoretical studies, the relationship between a local electric dipole moment and neighboring spins has been expressed by $\bm{p} \propto \bm{r}_{ij} \times (\bm{S}_i \times \bm{S}_j)$ (where $\bm{r}_{ij}$ is a vector connecting the two spins $\bm{S}_i$ and $\bm{S}_j$), based on the inverse Dzyaloshinskii-Moriya (DM) [4,5] and spin current mechanisms [6]. These theories can explain ferroelectric polarization in many multiferroics with cycloidal spin orderings, in which spins rotate in a plane parallel to the $\bm{k}$ vector, such as TbMnO$_3$ [7,8] and CoCr$_2$O$_4$ [9], because $\bm{r}_{ij} \perp \bm{S}_i \times \bm{S}_j$. An electric polarization component has, however, been identified even in the case of $\bm{r}_{ij} || \bm{S}_i \times \bm{S}_j$, in some multiferroics with proper screw ordering, in which the spins rotate in the plane perpendicular to the $\bm{k}$ vector. These include CuFeO$_2$ [10–16], CuCrO$_2$ [17–19], RbFe(MoO$_4$)$_2$ [20,21], and Cu$_3$Nb$_2$O$_8$ [22]. Arima subsequently succeeded in explaining the polarization induced by proper screw ordering in CuFeO$_2$ using the metal-ligand hybridization model with spin-orbit coupling [23,24]. Kaplan et al. [25] and Johnson et al. [21,22,26] extended the inverse DM mechanism to explain the emergence of polarization parallel to $\bm{S}_i \times \bm{S}_j$. Very recently, Tokunaga et al. reported that a polarization component orthogonal to the trigonal $c$ axis exists in the cycloidal phase of BiFeO$_3$ [27], which can also be explained by this mechanism [25].

α-NaFeO$_2$ is an $ABO_2$-type triangular lattice antiferromagnet [28], which has a rock-salt type crystal structure with the space group $R\bar{3}m$ [29] [Fig. 1(a)]. The structure of this material is similar to that of delafossite minerals such as CuFeO$_2$, apart from the oxygen coordination of the nonmagnetic $A^+$ ion, which is octahedrally surrounded by six $O^{2-}$ ions in α-NaFeO$_2$, as opposed to the linear coordination in the delafossites. In spite of the similar crystal structures of these compounds, especially the octahedrally coordinated magnetic Fe$^{3+}$ ions, the nearest-neighbor (NN) exchange interactions are completely different: ferromagnetic in α-NaFeO$_2$ [29,30] and antiferromagnetic in the delafossites [31,32]. However, a degree of frustration remains in α-NaFeO$_2$ due to the second NN antiferromagnetic interactions in the triangular lattice plane. In previous powder studies of α-NaFeO$_2$, successive magnetic phase transitions were observed, including an incommensurate spin-density-wave (SDW) phase (ICM1) over the range of $5.5 \leqslant T \leqslant 11$ K and a commensurate (CM) collinear phase below $T = 5.5$ K [33,34]. Terada *et al.* reported another incommensurate phase below 7.5 K, mixing with ICM1 and CM phases, accompanied by the onset of ferroelectric polarization [34]. The application of an external magnetic field increases the volume fraction of the incommensurate phase in the vicinity of 3 T, which is concomitant with a large increase in the ferroelectric polarization. Although powder diffraction data predict spiral ordering with an incommensurate propagation vector at a general point, $\boldsymbol{k} = (\alpha, \beta, \gamma)$, the relationship between the direction of the magnetic field and various phase transitions including magnetic orderings has not been clarified due to the lack of availability of single crystals. In the present study, we succeeded in growing single crystals of $\alpha$-NaFeO$_2$ using a hydrothermal synthesis method. Employing magnetization, specific heat, dielectric permittivity, and pyroelectric current measurements, as well as neutron diffraction experiments, we investigated the magnetic and dielectric properties of the multiferroic triangular lattice antiferromagnet $\alpha$-NaFeO$_2$.

*TERADA.Noriki@nims.go.jp
†hirohiko@phys.chuo-u.ac.jp

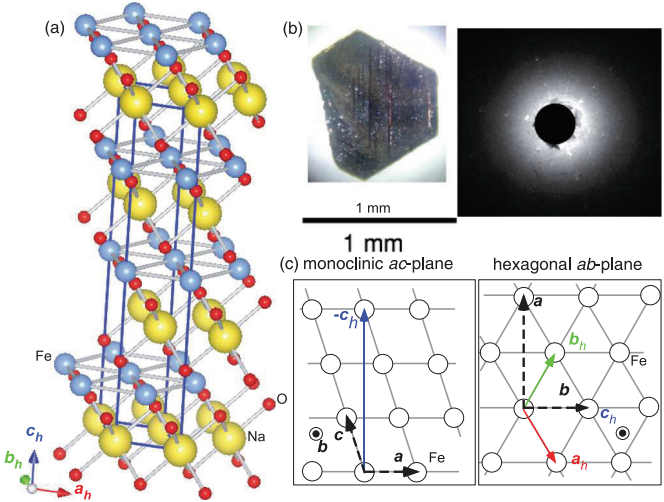

FIG. 1. (a) The crystal structure of $\alpha$-NaFeO$_2$. Large, medium, and small spheres denote Na$^+$, Fe$^{3+}$, and O$^{2-}$ ions, respectively. The Fe$^{3+}$ ions form triangle-lattice planes perpendicular to the $c$ axis. (b) A micrograph of $\alpha$-NaFeO$_2$ (in which the hexagonal-platelet shape reflects the crystal structure) and an x-ray Laue pattern (hexagonal $c$ direction) generated by the crystal. (c) The relationship between the hexagonal and monoclinic basis vectors [$\boldsymbol{a} = \boldsymbol{b}_h - \boldsymbol{a}_h$, $\boldsymbol{b} = \boldsymbol{a}_h + \boldsymbol{b}_h$, and $\boldsymbol{c} = \frac{1}{3}(\boldsymbol{a}_h - \boldsymbol{b}_h - \boldsymbol{c}_h)$, where Fe is at the origin] [34]. Solid and dotted lines denote monoclinic and hexagonal unit cells, respectively.

## II. EXPERIMENTAL DETAILS

Single crystals of $\alpha$-NaFeO$_2$ were grown using a hydrothermal method. In this process, the starting materials, FeO and NaOH, were sealed in a silver capsule with a small amount of H$_2$O. This mixture was kept at 650 °C and 150 MPa for one day. After the reaction, hexagonal pellets with a typical thickness of 0.5 mm, as shown in Fig. 1(b), were obtained. X-ray diffraction data confirmed that these were single crystals of $\alpha$-NaFeO$_2$. We confirmed that there is no magnetic impurity with a ferromagnetic component, such as $\beta$-NaFeO$_2$ and Fe$_2$O$_3$, by magnetization measurements on the single crystal.

Magnetization below 6.5 T was measured using a magnetic property measurement system (Quantum Design, MPMS-XL). Magnetization at higher magnetic fields up to 25 T was measured using a pulsed magnet at the Institute for Solid State Physics. Specific heat, dielectric permittivity, and pyroelectric current measurements were performed with a physical properties measurement system (Quantum Design, PPMS). The dielectric permittivity and pyroelectric current were determined using an LCR meter (Agilent, E4980A and NF, ZM2372) and an electrometer (Keithley, 6517B), respectively. A frequency of 100 kHz was employed for the dielectric permittivity measurements. During pyroelectric current measurements, the sample was first cooled in an 800–2000 kV/m poling electric field, after which the pyroelectric current was recorded on warming in zero electric field. Integrating the current with respect to time gave the dielectric polarization. We confirmed that the sign of the dielectric polarization was reversed when reversing the poling electric field. In these bulk measurements, we applied magnetic fields along one of three equivalent $\langle 110 \rangle$ directions in the case of $B_{ab}$ and along the $c$ axis in the case of $B_c$. Single-crystal neutron diffraction measurements were carried out using the WISH cold neutron time-of-flight diffractometer [35] at the ISIS Facility of the Rutherford Appleton Laboratory (UK), applying magnetic fields up to 13.4 T. The single crystal was mounted on a vertical field superconducting cryomagnet so that the magnetic field was applied along the monoclinic $b$ axis (hexagonal $\langle 110 \rangle$ axis) in the first experiment and along the monoclinic $c^*$ axis (hexagonal $c$ axis) in the second experiment. Hereafter, we use monoclinic notation unless otherwise specified. The relationship between the monoclinic and hexagonal bases is illustrated in Fig. 1(c). Crystal and magnetic structure refinements were performed using the FULLPROF program [36].

## III. RESULTS

### A. Bulk measurements

#### 1. Magnetization

Figure 2(a) shows the temperature dependences of the magnetic susceptibilities of $\alpha$-NaFeO$_2$ under various magnetic fields in the $ab$ plane ($B_{ab}$) and along the $c^*$ axis ($B_c$) (the same direction as the hexagonal $c$ axis). These data clearly show successive magnetic transitions. Under a 0.1 T magnetic field, the $\chi$ versus $T$ curve exhibits a bend at $T = 9$ K and a jump at $T = 5$ K. This behavior qualitatively reproduces the powder data reported by McQueen *et al.* [33] and Terada *et al.* [34]. In the present study, the transition at $T = 5$ K is very sharp, demonstrating the extremely high homogeneity of the crystal. The transition temperatures seen here are slightly different from those in the previous studies; McQueen *et al.* reported two magnetic transitions at $T = 11$ and 5 K, while Terada *et al.* found transitions at $T = 11$, 7.5, and 5.5 K. The decrease in the magnetic susceptibility below 9 K is more pronounced in $B_{ab}$. The steep drop in the susceptibility at $T = 5$ K is also more evident in $B_{ab}$. These results are consistent with neutron data, which indicate that the spins are oriented parallel to the $b$ axis in the lowest-temperature phase [33,34]. The transition temperatures depend strongly on the direction and the magnitude of the magnetic field. The lower-temperature transition with a steep drop disappears above $B_{ab} = 3.5$ T and $B_c = 5.5$ T. At $B_{ab} = 2.7$ T, the susceptibility increases with decreasing temperature below 4.8 K, while at $B_c = 5.0$ T a small drop in the susceptibility is observed at 3.8 K.

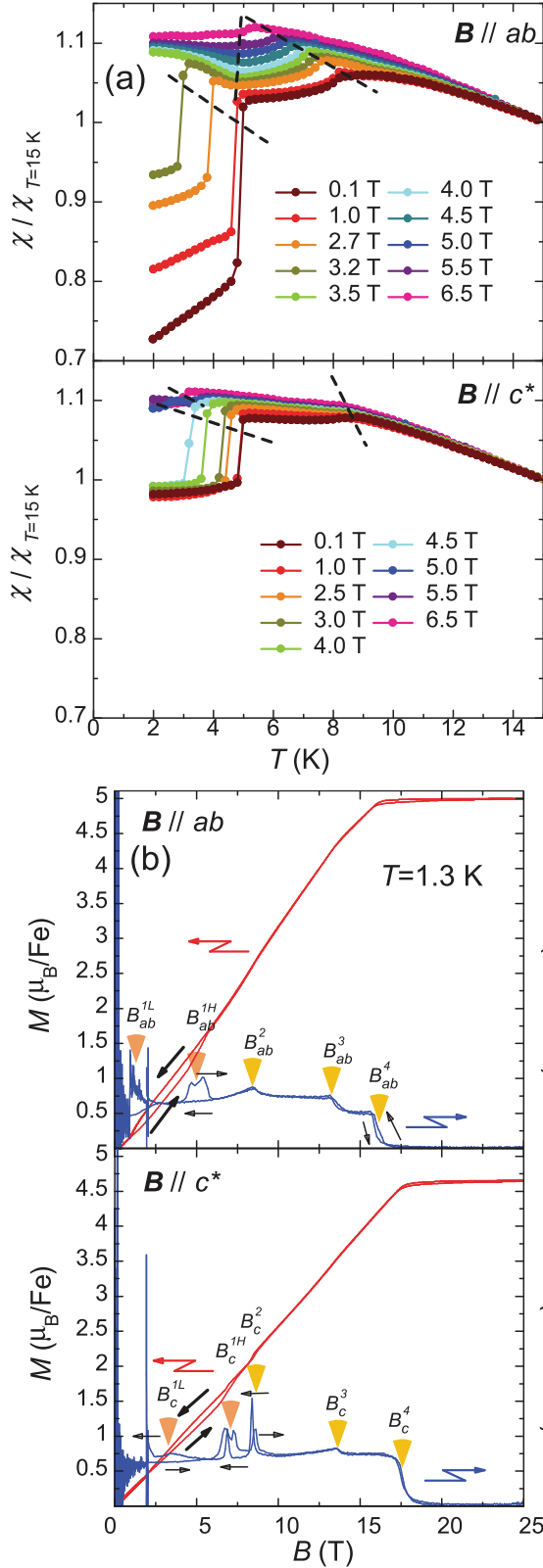

FIG. 2. (a) Temperature dependence of magnetization measured in a magnetic field up to 6.5 T applied along (top) the $b$ axis and (bottom) the $c^*$ axis. The dotted lines denote variations of the phase-transition temperatures. (b) Magnetization processes and the derivative of magnetization with respect to the magnetic field ($dM/dB$) along (top) the $b$ axis and (bottom) the $c^*$ axis. The triangular symbols denote peaks in the $dM/dB$ plots.

The magnetization curves measured for $\alpha$-NaFeO$_2$ using a pulsed magnetic field demonstrate multistep phase transitions, as shown in Fig. 2(b). In $B_{ab}$, significant hysteresis behavior is observed over a range of magnetic fields between 1.5 T ($B_{ab}^{1L}$) and 5 T ($B_{ab}^{1H}$), indicating a spin-flop-like first-order phase transition in this range. Considering the steep change in the $M$-$T$ curve above 3.3 T, as noted above, we can infer that the phase transition occurs around 3.3 T ($\equiv B_{ab}^1$), albeit with significant hysteresis. Similar behavior is observed in $B_c$ for 3 T ($B_c^{1L}$) $\leqslant B_c \leqslant$ 7 T ($B_c^{1L}$), corresponding to a first-order phase transition in the vicinity of 5 T ($\equiv B_c^1$). Above the first field-induced phase transitions, $dM/dB$ shows obvious peak anomalies at 8.5 T ($B_{ab}^2$), 13.6 T ($B_{ab}^3$), 16.0 T ($B_{ab}^4$), 8.5 T ($B_c^2$), 13.5 T ($B_c^3$), and 17.5 T ($B_c^4$). The phase transition at $B_c^2$ exhibits hysteresis indicative of a first-order transition, while those at $B_{ab}^2$, $B_{ab}^3$, $B_{ab}^4$, $B_c^3$, and $B_c^4$ are without hysteresis, suggesting second-order transitions. The saturation magnetization is approximately 5 $\mu_B$ per Fe atom, which is consistent with the expected value for Fe$^{3+}$ ions in the high-spin state.

### 2. Specific heat

Figure 3 presents the specific-heat curves of $\alpha$-NaFeO$_2$ acquired under various $B_{ab}$ and $B_c$ up to 14 T. Three-step phase transitions are clearly evident at $T = 8.5$ K ($T_{N1}$), 8.0 K ($T_{N2}$), and 4.8 K ($T_{N3}$) in zero field, whereas the magnetization measurements did not indicate a transition at 8 K, as shown in Fig. 2(a). The transitions at $T_{N2}$ and $T_{N3}$, denoted by triangles in Fig. 3, are accompanied by latent heat, indicating first-order transitions. These can be distinguished from the second-order transition at $T_{N1}$, which does not exhibit latent heat.

In $B_{ab}$, the transitions at $T_{N2}$ and $T_{N3}$ disappear above $B_{ab} = 4.0$ and 8.0 T, respectively. Instead, another second-order transition appears at $T = 4.7$ K, $B_{ab} = 3.0$ T, and gradually disappears above $B_{ab} = 7.0$ T. Fields greater than $B_{ab} = 5.0$ T induced another second-order transition at $T = 4.5$ K, which also gradually decreased with increasing $B_{ab}$ up to $B_{ab} = 13$ T. $B_c = 4$ T splits the transition at $T_{N3}$ into two first-order transitions at $T = 4.6$ and 4.0 K. The lower-temperature transition fades above $B_c = 5.0$ T, while the higher-temperature one remains up to $B_c = 13.0$ T. The transition at $T_{N2}$ at $B = 0$ T also persists up to the maximum $B_c$ applied. Some additional small peaks were evident above the highest phase-transition temperatures at $T = 5.9$ K and $B_{ab} = 13$ T, $T = 5.2$ K and $B_{ab} = 14$ T, $T = 8.2$ K and $B_c = 8$ T, and $T = 8.1$ K and $B_c = 9$ T, as indicated by the double arrows in Fig. 3. The origin of these peaks was not determined in this study.

### 3. Dielectric permittivity and polarization

Figure 4 summarizes the temperature dependences of the dielectric permittivity values of $\alpha$-NaFeO$_2$ under $B_{ab}$ and $B_c$. In zero field, a weak anomaly with thermal hysteresis, corresponding to a magnetic transition, is observed at $T = 5$ K. The shape of this anomaly is extremely asymmetric; the slope on the low-temperature side is much steeper than that on the high-temperature side. Such behavior is characteristic of a first-order transition. Above $B_{ab} = 3.0$ T or $B_c = 5.0$ T, the peak becomes very pronounced and exhibits divergent behavior on both sides of the transition temperature. This indicates that another dielectric phase is induced by the magnetic fields. In addition, small anomalies, indicated by triangles, appear under magnetic fields higher than $B_c = 4$ T, while these anomalies do not appear when an electric field is applied along the $c^*$ axis. The anomalies observed in the dielectric permittivity are also detected in the specific-heat measurements, apart from those at $T = 3.5, 3.2$, and $2.8$ K, at $B_c = 7.0$ T, due to the low-temperature limit associated with the present specific-heat measurements. We should mention here the possibility of artificial magnetoelectric coupling, which gives rise to a strong frequency dependence of dielectric permittivity described in previous papers [37,38]. We checked the frequency dependence of dielectric permittivity, which does not show such a behavior at the low-temperature range. We can therefore exclude the possibility.

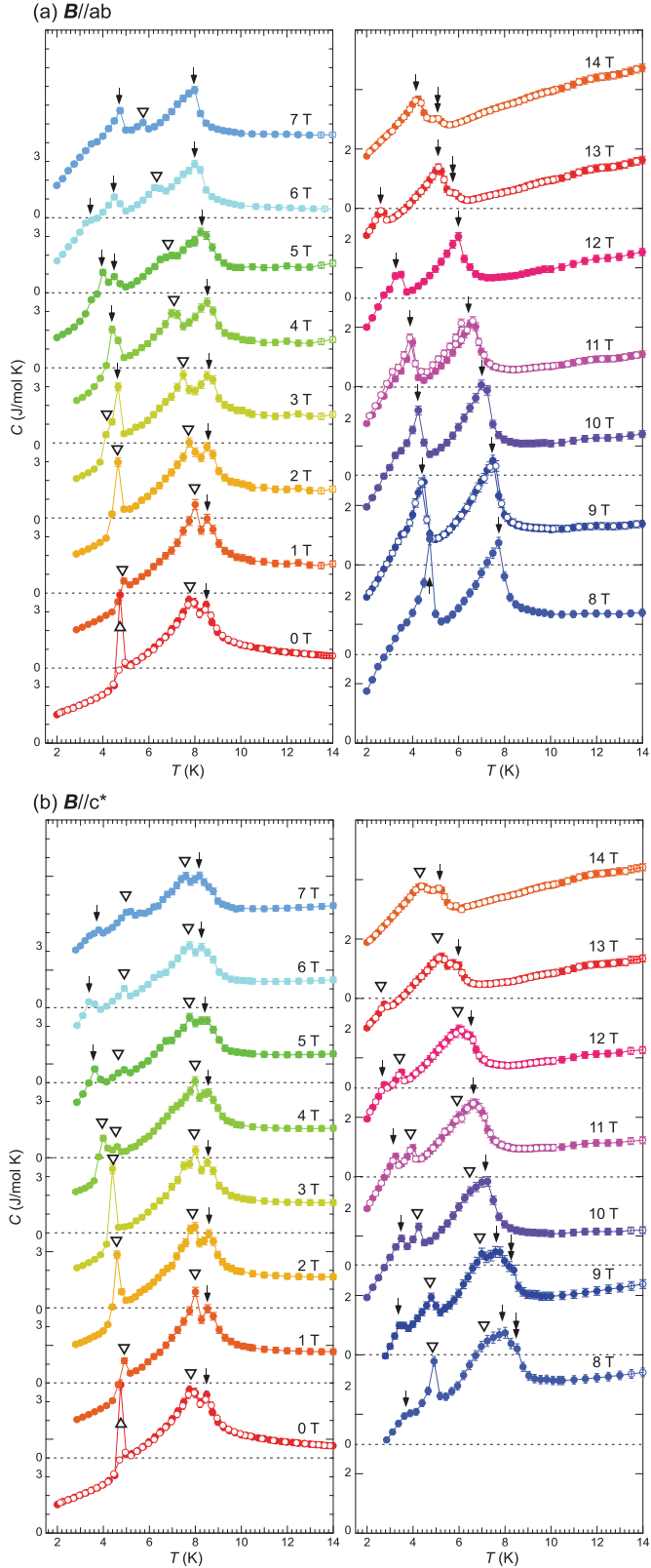

FIG. 3. Temperature dependences of the specific-heat values of $\alpha$-NaFeO$_2$ under various magnetic fields (a) along the $ab$ plane and (b) along the $c$ axis. These data were acquired while increasing the temperature. The triangles and arrows denote first- and second-order phase transitions, respectively.

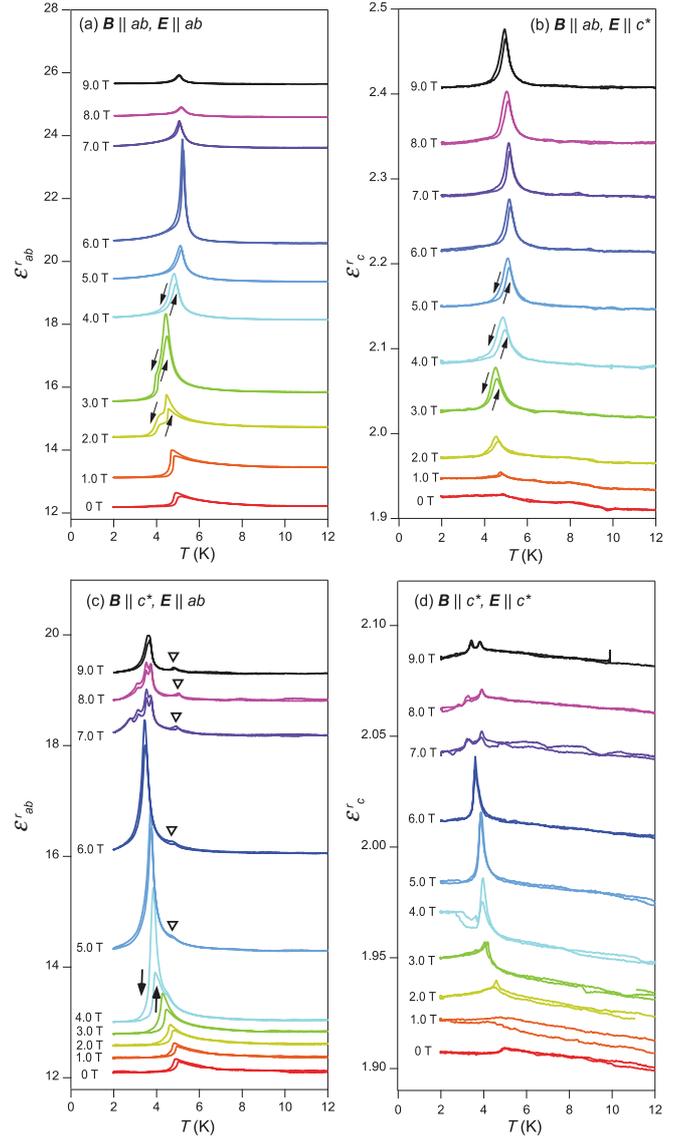

FIG. 4. Temperature dependences of the relative dielectric permittivity values under various magnetic fields. The triangles denote small anomalies, and the arrows indicate the direction of the temperature change. The directions of the applied magnetic and electric fields were as follows: (a) $\boldsymbol{B} \parallel ab$ and $\boldsymbol{E} \parallel ab$, (b) $\boldsymbol{B} \parallel ab$ and $\boldsymbol{E} \parallel c^*$, (c) $\boldsymbol{B} \parallel c^*$ and $\boldsymbol{E} \parallel ab$, and (d) $\boldsymbol{B} \parallel c^*$ and $\boldsymbol{E} \parallel c^*$.

Spontaneous electric polarizations were induced above $B_{ab} = 3$ T and $B_c = 5$ T, as shown in Fig. 5. In the case of $B_{ab} > 3$ T, electric polarizations parallel to the $ab$ plane ($P_{ab}$) and the $c^*$ axis ($P_c$) were observed simultaneously, indicating the emergence of polarization that was not confined to any highly symmetric direction (general direction), as shown in Figs. 5(a) and 5(b). It should be noted that we were unable to determine the direction of polarization in the $ab$ plane for $P_{ab}$ because three equivalent magnetic domains were present at 120° intervals. The observed polarization values in $B_{ab}$, $P_{ab} \simeq 50$ $\mu$C/m$^2$, and $P_c \simeq 10$ $\mu$C/m$^2$ are comparable to those reported for typical type II multiferroics [2,3]. In contrast, in the case of $B_c > 5$ T and $3.8 \lesssim T \lesssim 5$ K, only $P_{ab}$ is observed without $P_c$, indicating that the polarization was confined to the $ab$ plane in this temperature region (defined as phase VI$_c$ below), as shown in Figs. 5(c) and 5(d). Below $T = 3.8$ K in $B_c$, both $P_c$ and $P_{ab}$ are observed as $P_{ab} \simeq 60$ $\mu$C/m$^2$ and $P_c \simeq 14$ $\mu$C/m$^2$. The temperature dependences of $P_{ab}$ and $P_c$ in $B_c$ show complicated stepwise changes appearing at the same temperatures at which dielectric permittivity and specific heat exhibit anomalies.

### B. Magnetoelectric phase diagrams

Summarizing the temperatures and magnetic fields at which anomalies were found in the bulk measurements described above, we obtained magnetic and dielectric phase diagrams as functions of temperature and $B_{ab}$ and $B_c$, as illustrated in Figs. 6(a) and 6(b). In zero magnetic field, three magnetic phase transitions occurred, including one second-order phase transition at 8.5 K($T_{N1}$) and two first-order transitions at 8 K($T_{N2}$) and 4.8 K($T_{N3}$). Herein, we define the magnetic phases in the temperatures ranges $T_{N2} \leqslant T \leqslant T_{N1}$, $T_{N3} \leqslant T \leqslant T_{N2}$, and $T \leqslant T_{N3}$ as phases I, II, and III. Applying $B_{ab}$ and $B_c$, we found significant and strong competition of the phase boundaries in the phase diagrams. This behavior indicates that $\alpha$-NaFeO$_2$ possesses strong spin frustration in addition to ferromagnetic NN exchange interactions.

Below $T_{N3}$, the application of $B_{ab}$ or $B_c$ induces ferroelectric phases in the field ranges of 3.3 T ($B^1_{ab}$) $\leqslant B_{ab} \leqslant$ 8.5 T ($B^2_{ab}$), $B^2_{ab} \leqslant B_{ab} \leqslant$ 13.6 T ($B^3_{ab}$), 5.0 T ($B^1_c$) $\leqslant B_c \leqslant$ 8.5 T ($B^2_c$), and $B^2_c \leqslant B_c \leqslant$ 13.5 T ($B^3_c$), indicated by IV$_{ab}$, V$_{ab}$, IV$_c$, and V$_c$ in Figs. 6(a) and 6(b). The ferroelectric polarizations in these phases point to a general direction, possessing both in-plane (hexagonal $ab$ plane) and $c^*$ (hexagonal $c$) polarization components, as noted in the previous section. In addition, another ferroelectric phase, defined as VI$_c$, appeared in the intermediate temperature region around 4 K and $5 \leqslant B_c \leqslant 13$ T. The ferroelectric polarization in the VI$_c$ phase was confined to the $ab$ plane, as shown in Fig. 5(b).

### C. Neutron diffraction
#### 1. Zero magnetic field

The experimental configuration in the case of a zero magnetic field was such that the monoclinic $b$ axis for one reflections with different $k$ vectors, $(q_a, q_b, q_c)$ ($\equiv k_{\mathrm{ICM2}}$) with $q_a \simeq 0.03$, $q_b \simeq 0.24$, and $q_c \simeq 0.49$ below $T_{N2}$ in phase II, as clearly seen in the contour map obtained at 6 K and shown in Fig. 8(b). While $k = (0, q, \frac{1}{2})$ in phase I is a line of symmetry and maintains the monoclinic symmetry with three monoclinic domains, the phase II $k$ vector, $(q_a, q_b, q_c)$ of the three magnetic domains (the hexagonal $\langle 100 \rangle$ axis) was vertical, a scenario that was identical to the magnetic field setup in the $ab$ plane illustrated in Fig. 7(a). These three domains were separated from the higher symmetric rhombohedral lattice with threefold symmetry along the hexagonal $c$ axis in the paramagnetic phase.

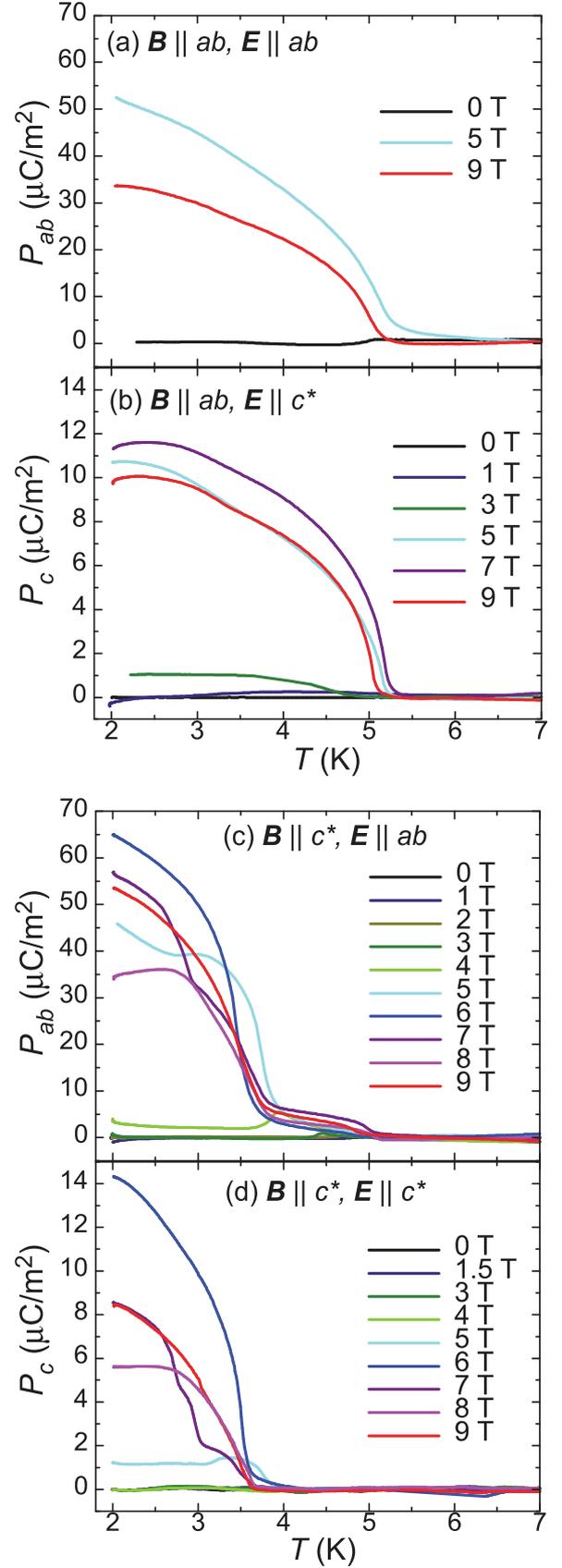

FIG. 5. Temperature dependences of the dielectric polarizations under the various applied magnetic fields (a) within the $a$-$b$ plane and (b) along the $c$ axis. The upper (lower) frame shows the out-of-plane (in-plane) component of the polarization.

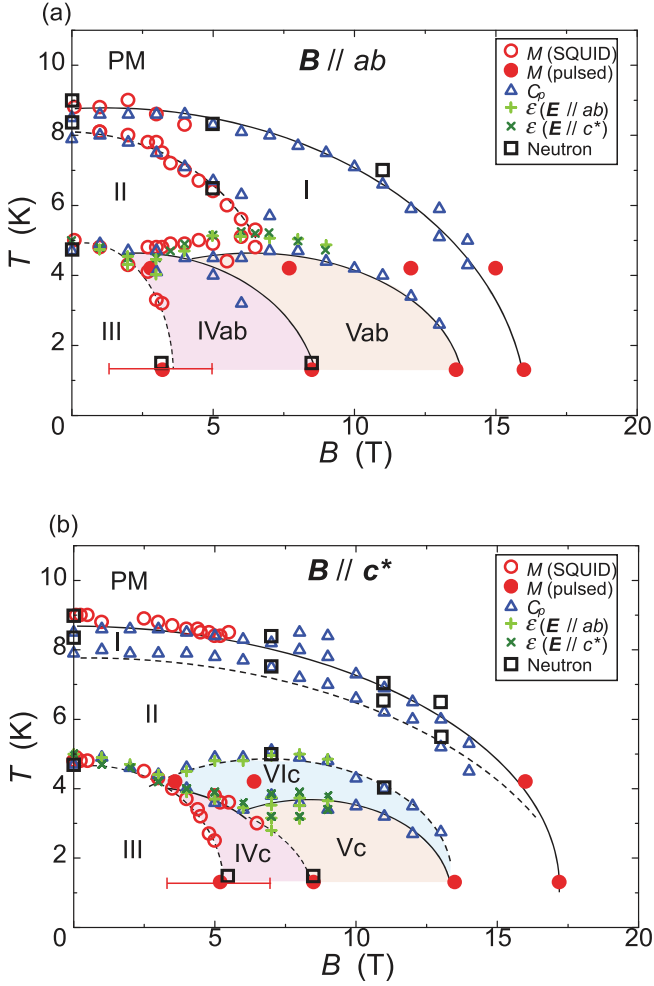

FIG. 6. Magnetic phase diagrams for $\alpha$-NaFeO$_2$ based on magnetization data acquired under a static magnetic field (open circles) and a pulsed magnetic field (filled circles), together with the results of specific heat (triangles), dielectric permittivity (crosses), and neutron diffraction measurements (squares). Horizontal bars denote the hysteresis region observed during magnetization measurements with a pulsed magnetic field. Broken and solid lines indicate the first- and second-order phase boundaries. The colored areas show the ferroelectric phases.

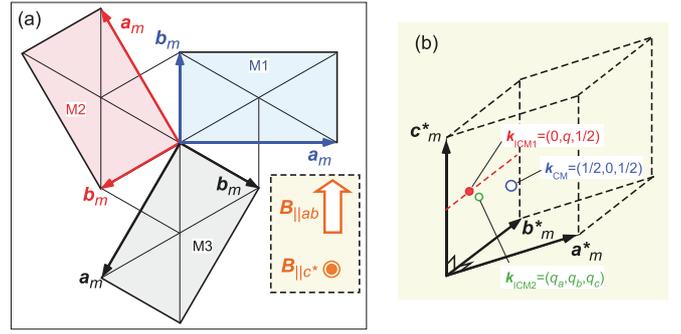

FIG. 7. (a) Schematic illustration of the relationship between the three monoclinic domains, including their basis vectors and magnetic field directions, in two different experimental configurations. (b) The reciprocal lattice in the monoclinic setting, including three different types of $k$ vectors for the magnetic phases of $\alpha$-NaFeO$_2$: $k_{\mathrm{CM}} = (\frac{1}{2}, 0, \frac{1}{2})$ for phase III, $k_{\mathrm{ICM1}} = (0, q, \frac{1}{2})$ for phase I, and $k_{\mathrm{ICM2}} = (q_a, q_b, q_c)$ for phases II, IV$_{ab}$, V$_{ab}$, IV$_c$, V$_c$, and VI$_c$.

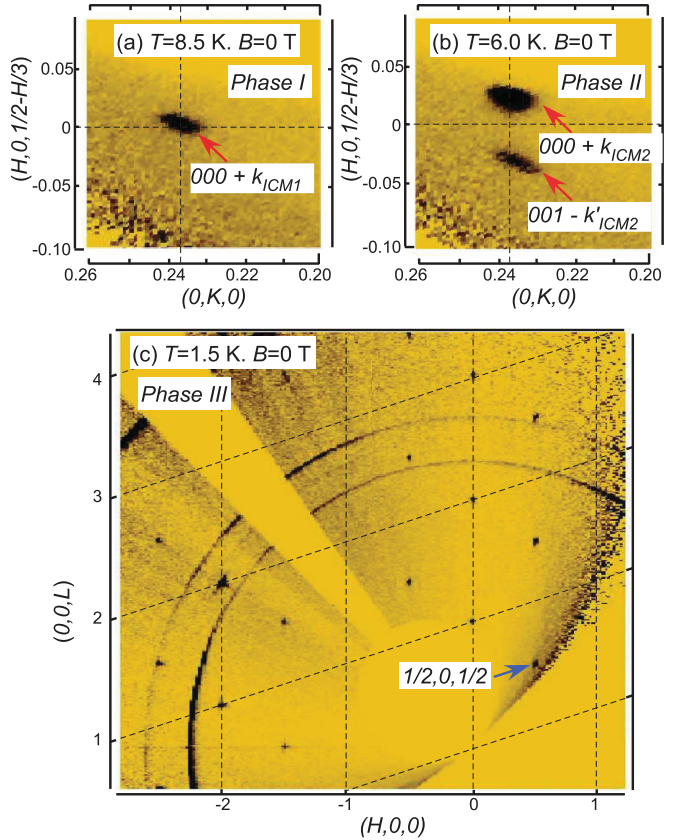

FIG. 8. Contour maps of neutron intensity at (a) 8.5 K, (b) 6.0 K, and (c) 1.5 K in zero magnetic field. The $(H, K, \frac{1}{2} - \frac{H}{3})$ reciprocal-lattice plane for domain $M2$ is shown in (a) and (b), and the $(H, 0, L)$ plane for domain $M1$ is shown in (c).

In phase I, a magnetic reflection assigned to $k = (0, q, \frac{1}{2})$ ($\equiv k_{\mathrm{ICM1}}$) with $q \simeq 0.24$ was observed at 8.5 K in the magnetic domain $M2$. The contour map for neutron intensity at 8.5 K is shown in Fig. 8(a). Here, the $k$ vector is consistent with that seen in previous powder studies [33,34]. We observed other magnetic reflections belonging to domains $M1$ and $M2$ in zero magnetic field, while reflections in domain $M3$ were not measurable due to geometric restrictions in these experiments. Upon decreasing the temperature from phase I, the incommensurate reflection was separated into two is a general point of the Brillouin zone and reduces the symmetry to triclinic, separating each monoclinic domain into two triclinic domains. Therefore, the two reflections in the contour map at 6.0 K [Fig. 8(b)] can be assigned to $k = (q_a, q_b, q_c)$ and $k' = (q_a, -q_b, q_c)$, equivalent to $000 + k$ and $001 - k'$. The incommensurabilities of the $a$ and $c$ components of the general point $k$ vector are significantly affected by temperature, as shown in Figs. 9(d) and 9(f). Below 4.8 K ($T_{N3}$), the incommensurate reflections disappear. Instead, commensurate reflections for which $k = (0.5, 0, 0.5)$ ($\equiv k_{CM}$) appear in phase III, as is evident in Figs. 8(c) and 9(c). This $k$ vector is consistent with reports resulting from previous powder studies [29,33,34].

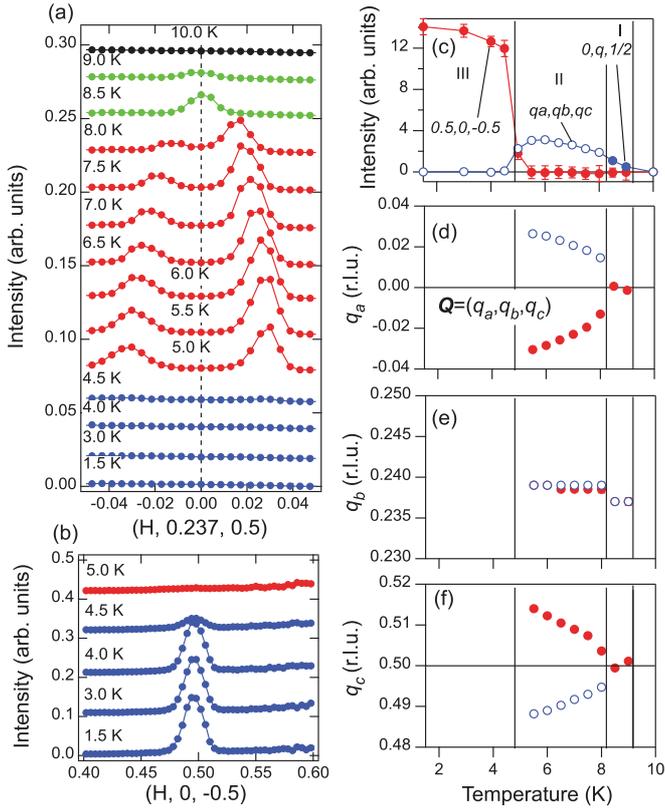

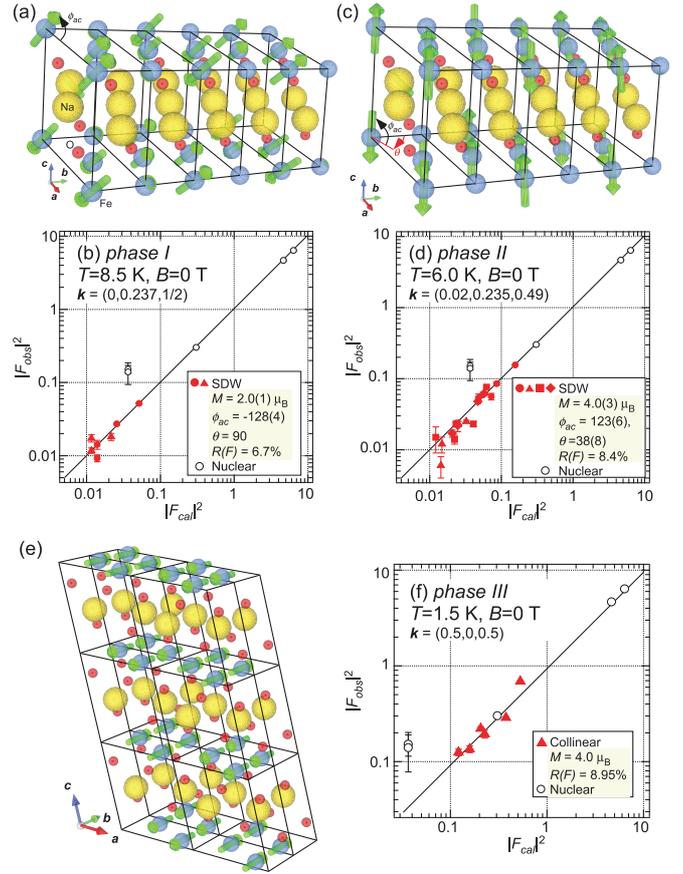

FIG. 9. Neutron diffraction line profiles along (a) $(H, 0.237, 0.5)$ (domain $M2$) and (b) $(H, 0, -0.5)$ (domain $M1$) for typical temperatures in zero magnetic field. The temperature dependence of (c) integrated intensity, and the incommensurability (d) $q_a$, (e) $q_b$, and (f) $q_c$ of the $k$ vector in zero field. In (d)–(f) the open and closed symbols denote the peak positions of $\mathbf{Q} = (q_a, q_b, q_c)$ and $\mathbf{Q} = (-q_a, q_b, 1 - q_c)$, respectively.

FIG. 10. Schematic drawings of the spin-density-wave structures in (a) phase I and (c) phase II, and (e) the collinear structure in phase III. Results of crystal and magnetic structure refinements for (b) phase I, (d) phase II, and (f) phase III. Solid triangles and circles denote data for magnetic reflections of domains $M1$ and $M2$, respectively.

To determine the magnetic structure in each phase, we refined the magnetic structure parameters based on intensity data acquired at typical temperatures. It should be noted that, in this analysis, we fixed the scale factor, which depends on the domain population, by assuming that a magnetic moment of 4.0 $\mu_B$ was stabilized at the lowest temperature, as expected based on our previous powder study [34]. In phase I, an SDW structure with collinear spins canting in the $ac$ plane from the $a$ axis, $\phi_{ac} = -128(4)°$, gave the best agreement with the experimental data obtained at 8.5 K, as shown in Figs. 10(a) and 10(b). For phase I with $k = (0, q, \frac{1}{2})$, there are two possible time-odd irreducible representations (IRs) of $R\bar{3}m1'$: $mY_1$ and $mY_2$ (in ISODISTORT notation [39,40]). The refined SDW in phase I is represented by the order parameter direction (OPD) $P(a, 0; 0, 0; 0, 0)$ in $mY_1$ IR space, restricting the spin direction in the $ac$ plane, which in turn reduces the symmetry to the monoclinic (3 + 1) superspace group $C2/m1'(0, \beta, \frac{1}{2})s0s$ ($\beta = q$) [39,40]. The resulting magnetic point group in phase I is nonpolar $2/m1'$, which agrees with the absence of electric polarization.

For phase II, the $k$ vector is a general point of symmetry, $k = (0.03, 0.239, 0.49)$, at 6.0 K, and this reduces the superspace group symmetry to triclinic: either $P\bar{1}1'(\alpha, \beta, \gamma)0s$, $P11'(\alpha, \beta, \gamma)0s$, or $R\bar{3}m1'$ [39,40]. The refinement gives the best results for the data at 6.0 K with a collinear SDW structure having moments represented by canting angles $\phi_{ac} = 98(10)°$ and $\theta = 124(4)°$, where $\theta$ is the tilt angle from the $b$ axis, as shown in Figs. 10(c) and 10(d). This SDW structure belongs to the superspace group $P\bar{1}1'(\alpha, \beta, \gamma)0s$ and the nonpolar magnetic point group $\bar{1}1'$. Since there is a single IR, $mGP_1$ [with the OPD $P(a, 0; 0, 0; 0, 0; 0, 0; 0, 0)$], for which the moment direction is not restricted, the refined spin direction is not confined to any symmetric direction, in contrast to phase I.

The magnetic structure for phase III in the ground state in zero magnetic field is consistent with results from previous powder studies [29,33,34], as shown in Figs. 10(e) and 10(f). Here, collinear spins pointing along the $b$ axis align as ↑↑↓↓ along the $a$ axis. The magnetic space group is also consistent with the previously reported $P_a2_1/m$ (in BNS notation) [34].

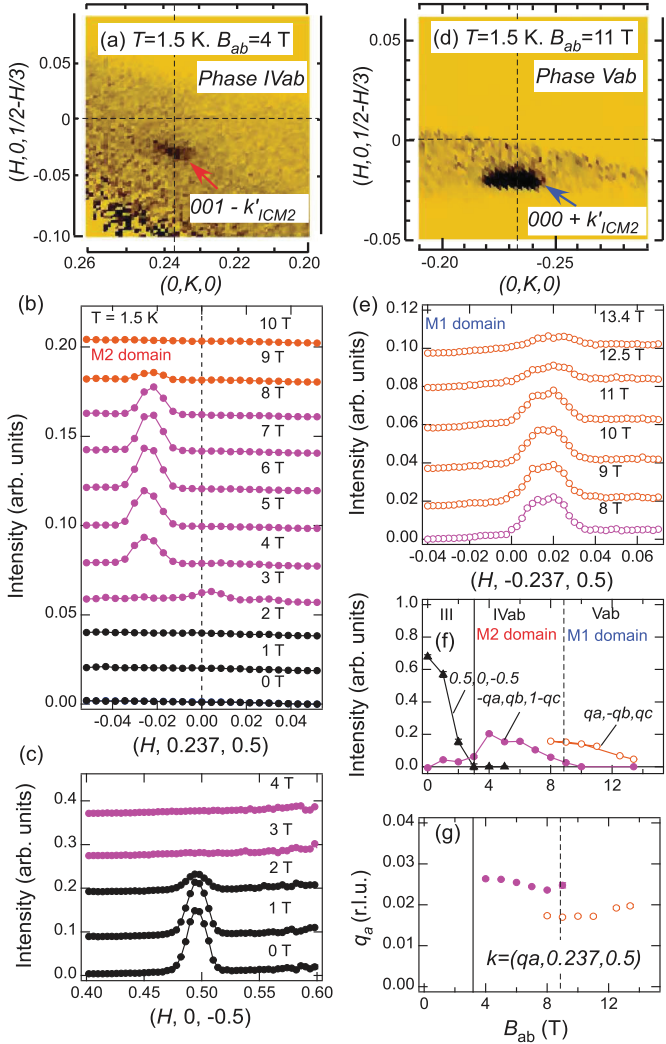

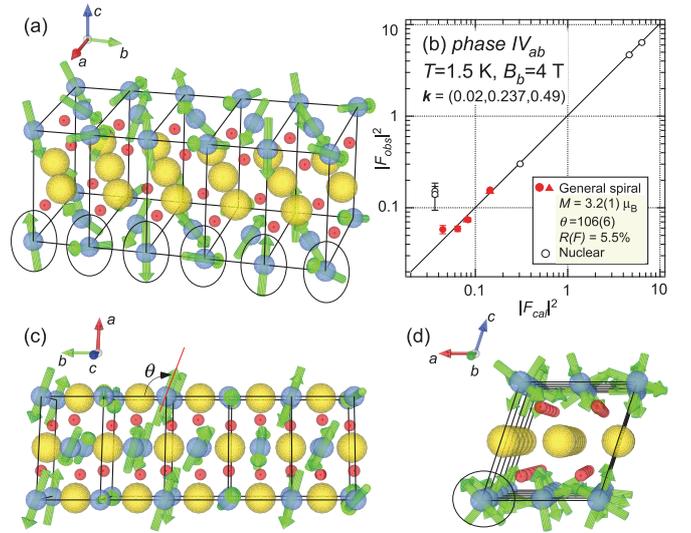

FIG. 11. Contour maps of neutron intensity measured at (a) $T = 1.5$ K and $B_{ab} = 4$ T (phase IV$_{ab}$) and (d) $T = 1.5$ K and $B_{ab} = 10$ T (phase V$_{ab}$). Neutron diffraction patterns at typical $B_{ab}$ along the (b) $(H,0.237,0.5)$, (c) $(H,0,-0.5)$, and (d) $(H,-0.237,0.5)$ lines. $B_{ab}$ dependences of (f) the integrated magnetic reflection intensity and (g) the $a$ component of $k_{ICM2}$. In (f) and (g) the closed and open circles denote data for the M1 and M2 domains, respectively.

#### 2. $B \parallel ab$

Magnetic phase transitions were clearly observed in the field dependence of neutron diffraction profiles under a magnetic field along the $ab$ plane at 1.5 K, as shown in Figs. 11(b), 11(c), and 11(e). The magnetic reflections assigned to commensurate $k_{CM}$ for phase III disappear at $B_{ab} = 3$ T, while incommensurate reflections associated with $k_{ICM2} = (q_a, q_b, q_c)$ appear [Figs. 11(b), 11(c), and 11(f)]. These data indicate a phase transition from phase III or IV$_{ab}$. As shown in Fig. 11(a), a reflection assigned to $001-k'_{ICM2}$ [$k'_{ICM2} = (q_a, -q_b, q_c)$] is observed in phase IV$_{ab}$; however, a reflection resulting from the other triclinic domain, $000+k_{ICM2}$, is not evident. This result suggests that one of the triclinic domains was arranged by the in-plane magnetic field, $B_{ab}$.

With further increases in $B_{ab}$, the $001-k'_{ICM2}$ reflection disappears at approximately $B_{ab} = 8.5$ T [Figs. 11(b) and 11(f)]. This reflection belonged to the monoclinic domain M2, as illustrated in Fig. 7(a). Conversely, an incommensurate reflection assigned to $000+k'_{ICM2}$ and belonging to another monoclinic domain (M1) remains even above $B_{ab} = 8.5$ T, as shown in Figs. 11(d) and 11(e). In addition, although the $k$-vector symmetry did not change at $B_{ab} = 8.5$ T, the $a$-component $k_{ICM2}$, $q_a$ shows a discontinuous change, as can be seen from Fig. 11(g). Thus, the domain rearrangement and the change in the $k$-vector component occurred simultaneously at $B_{ab} = 8.5$ T. Considering the peak anomaly in the $dM/dB$ plot at $B_{ab} = 8.5$ T, which shows no hysteresis upon increasing and decreasing the field, we can conclude that the magnetic phase transition is from phase IV$_{ab}$ to V$_{ab}$ at $B_{ab} = 8.5$ T. The intensity of the $000+k'_{ICM2}$ reflection for phase V$_{ab}$ decreases just above $B_{ab} = 13.5$ T (which was the highest field achievable experimentally), indicating another transition to phase I.

Magnetic structural determination for phase IV$_{ab}$ was carried out based on the intensity data acquired at $B_{ab} = 4$ T and 1.5 K. We obtained the best refinement with a spiral model expressed by the two-dimensional OPD $C(a,b;0,0;0,0;0,0;0,0)$ in the $mGP_1$ IR space [39,40]. The results of refinement and the spin model for phase IV$_{ab}$ are shown in Fig. 12. Since the spiral plane is tilted by $\theta = 106(6)°$ from the $b$ axis [Figs. 12(c) and 12(d)], the spiral structure has both proper screw and cycloid components. Hereafter, we refer to this structure as the "general spiral." It should also be noted that, in the case of a proper screw structure, the spins lie on the plane perpendicular to the $b$ axis ($\theta = 90°$), while the spiral plane of the cycloid structure is parallel to the $b$ axis. As well, note that, due to the slight incommensurability of the $a$ and $c$ components of the $\mathbf{k}$ vector, the period of the spiral modulation is very long along the $a$ and $c$ directions. This general spiral structure in phase IV$_{ab}$ reduces the symmetry to the superspace group $P11'(\alpha,\beta,\gamma)0s$ and the polar magnetic point group $11'$. This polar point group allows electric polarization in the general direction, which is in agreement with the results of polarization measurements. In the case of the V$_{ab}$ phase, we could not refine the magnetic structural parameters due to an insufficient quantity of reflections obtained with our experimental geometry.

FIG. 12. (a) Illustration of the general spiral structure of phase IV$_{ab}$ $\alpha$-NaFeO2. (c) and (d) Projections from the $c$ and $b$ axis. Here, $\theta$ is the angle between the spiral plane and the $b$ axis. (d) The relationship between the experimental and calculated structure factors. The refined parameters are also shown in the inset. Open and closed symbols denote nuclear and magnetic reflections, and differences in the symbols for the magnetic data correspond to different domains.

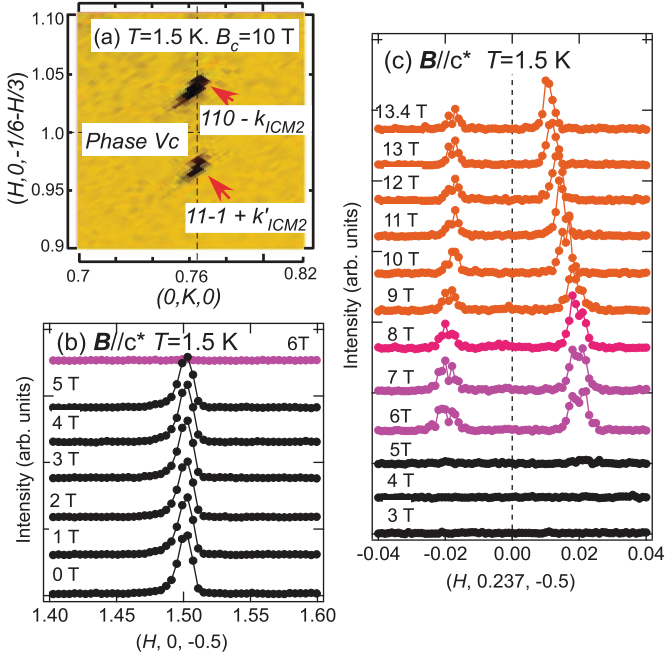

FIG. 13. (a) Contour map for neutron intensity at $B_c = 10$ T and $T = 1.5$ K in phase $V_c$, showing two peaks assigned as $(1,1,0) - \mathbf{k}_{ICM2}$ and $(1,1,\bar{1}) + \mathbf{k}'_{ICM2}$. The $B_c$ dependence of the neutron diffraction profiles along the (b) $(H,0,-0.5)$ and (c) $(H,0.237,-0.5)$ lines at $T = 1.5$ K.

### 3. $B \parallel c^*$

The application of a magnetic field along the $c^*$ axis (hexagonal $c$ axis) generated a phase transition at $B_c = 5.5$ T and $T = 1.5$ K. As shown in Figs. 13(b) and 14(a), the magnetic reflection for the ground state in zero field at $(0,5,0,-0.5)$ for phase III disappears. Instead, incommensurate peaks assigned to $\mathbf{k}_{ICM2} = (q_a, q_b, q_c)$ are induced in phase IV$_c$, as can be seen in Fig. 14(c). The $\mathbf{k}$ vector is a general point of symmetry similar to phases II, IV$_{ab}$, and V$_{ab}$. Since the $\mathbf{B}_c$ direction is parallel to the threefold axis in the parent space group, $R\bar{3}m$, rearrangement of the monoclinic domains [$M1$, $M2$, and $M3$ in Fig. 7(a)] was not observed. The triclinic domains in each monoclinic domain, $\mathbf{k}_{ICM2}$ and $\mathbf{k}'_{ICM2}$, also maintained finite populations, as demonstrated by the presence of the double peak profiles seen in Figs. 13(a) and 13(c). Above $B_c = 9$ T, the diffraction patterns remained unchanged, while the integrated intensity of the magnetic peak was slightly reduced, and one of the $\mathbf{k}$-vector components, $q_b$, also began to change with increasing $B_c$, as shown in Figs. 14(a) and 14(c). The other $\mathbf{k}$-vector components also exhibit a small amount of variation [Figs. 14(b) and 14(d)]. These anomalies correspond to the transition from phase IV$_c$ to V$_c$.

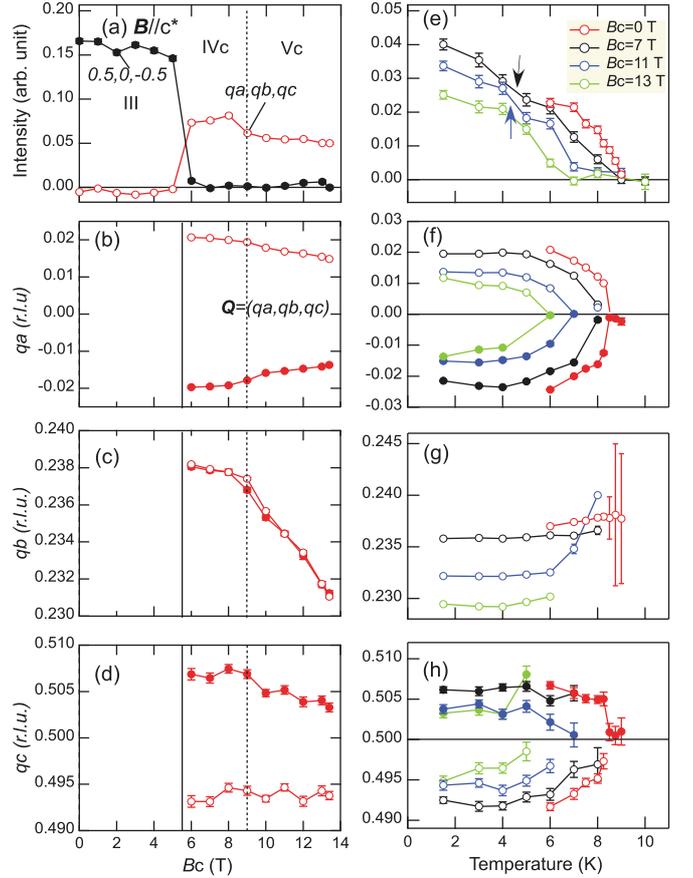

FIG. 14. Effects of the magnetic field parallel to the $c^*$ axis on the (a) integrated intensity, and the (b) $a$, (c) $b$, and (d) $c$ components of $Q$ for observed magnetic reflections at $T = 1.5$ K. Temperature variations of (e) the integrated intensity, and (f)–(h) of the three components of $Q$ at $B_c = 0, 7, 11$, and 13 T, respectively. The open and closed symbols in (b)–(d) and (f)–(h) denote the $Q = (q_a, q_b, q_c)$ and $Q = (-q_a, q_b, 1 - q_c)$ peak positions, respectively.

The phase transition from phase I to phase II under $B_c$ is clearly observed in the temperature dependence of the $\mathbf{k}$-vector components presented in Figs. 14(f), 14(g), and 14(h). As noted in the previous section, the $\mathbf{k}$-vector symmetry also transitioned from $\mathbf{k}_{ICM1} = (0, q, \frac{1}{2})$ to $\mathbf{k}_{ICM2} = (q_a, q_b, q_c)$ as a function of temperature under $B_c$, as demonstrated by the observation of separate magnetic Bragg peaks. With further decreases in temperature, the integrated intensity of the $q_a, q_b, q_c$ reflection began to increase, as indicated by the arrows in Fig. 14(e). In addition, whereas the $\mathbf{k}$-vector components changed significantly as a function of temperature in phase II, these components were constant below the phase transition temperature to phase VI$_c$. This phase transition is also concomitant with the onset of electric polarization perpendicular to the $c^*$ axis, as shown in Fig. 5(b). Although the integrated intensity plot exhibits a slight change in curvature in the vicinity of $T = 3$ K [Fig. 14(e)], we do not observe any significant anomalies in the temperature dependence of the magnetic reflection at the phase transition from VI$_c$ to IV$_c$ (V$_c$).

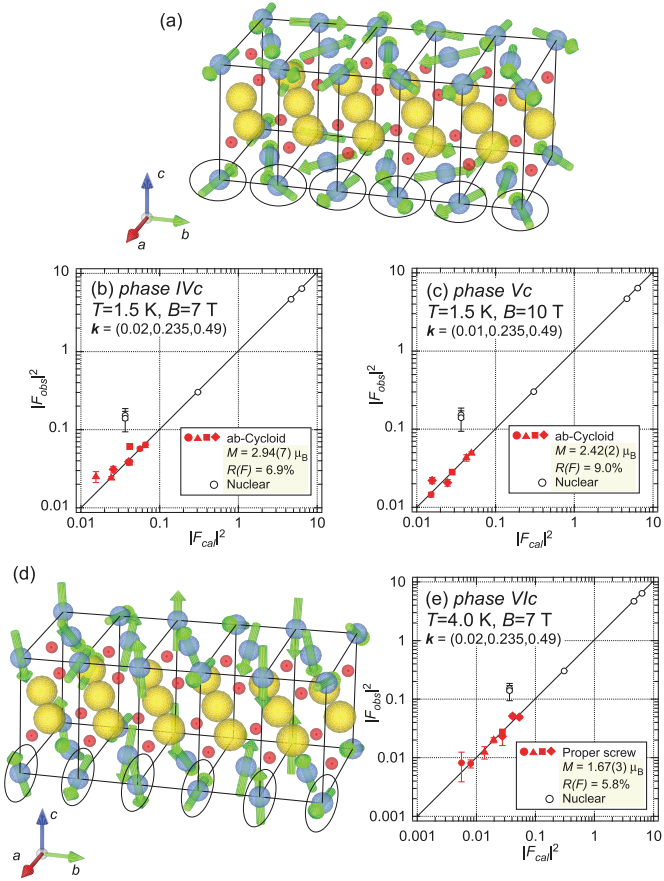

FIG. 15. Illustrations of the magnetic structures determined for the (a) $ab$ cycloid and (c) proper screw. Results of the refinement of data acquired at (b) $B_c = 7$ T (phase IV$_c$) and 10 T (phase V$_c$) at $T = 1.5$ K, and $B_c = 7$ T and $T = 4.0$ K (phase VI$_c$). The open and closed symbols denote nuclear and magnetic reflections, and the differences in the symbols for magnetic data correspond to different domains.

Magnetic structure refinements of the magnetic-field-induced phases were performed using data acquired at $B_c = 7$ T and $T = 1.5$ K (phase IV$_c$), $B_c = 10$ T and $T = 1.5$ K (phase V$_c$), and $B_c = 7$ T and $T = 4.0$ K (phase VI$_c$). For phases IV$_c$ and V$_c$, we succeeded in refining the data against an $ab$-cycloid structure model having $a$ and $b$ spin components, as illustrated in Fig. 15(a). The refinement results are provided in Figs. 15(b) and 15(c). As discussed in the previous section, there is one time-odd IR, $mGP_1$, when we have a $k$ vector with a general point of symmetry in the case of the parent space group of $R\bar{3}m1'$. The $ab$-cycloid structure is expressed by the OPD $C(a,b;0,0;0,0;0,0;0,0;0,0)$ in the $mGP_1$ IR space. The resultant superspace group is $P11'(\alpha,\beta,\gamma)0s$ (and the magnetic point group is $11'$), which is identical to that of phase IV$_{ab}$. The observed electric polarization in phases IV$_c$ and V$_c$ points in the general direction [Figs. 5(a) and 5(b)], which is in agreement with the magnetic symmetry.

In contrast, in the case of phase VI$_c$, we did not observe an electric polarization parallel to the $c^*$ axis but only in the $ab$ plane [Figs. 5(a) and 5(b)]. The magnetic structural determination was carried out with the data obtained at $B_c = 7$ T and $T = 4.0$ K for phase VI$_c$. Unlike phases IV$_c$ and V$_c$, the magnetic structure in phase VI$_c$ is a proper screw structure with spin components in the $ac$ plane, as shown in Figs. 15(d) and 15(e). Since the $k$ vector is on a general point in phase VI$_c$, the proper screw structure is also expressed by the OPD $C(a,b;0,0;0,0;0,0;0,0)$ in the $mGP_1$ IR space, and it has the superspace group $P11'(\alpha,\beta,\gamma)0s$. The magnetic symmetry does not confine the electric polarization to any specific direction, although the electric polarization observed in the experiment was only in the $ab$ plane in phase VI$_c$. This point is discussed in more detail below.

## IV. DISCUSSION

### A. Direction of polarization

At this point, we can discuss the relationship between the observed electric polarization and the magnetic ordering in each polar phase. Electric polarization was observed in phases IV$_{ab}$, V$_{ab}$, IV$_c$, V$_c$, and VI$_c$, each of which has different magnetic structure with the same type of $k$ vector, $k_{\text{ICM2}} = (q_a, q_b, q_c)$, and the same triclinic magnetic point group, $11'$. The first four phases exhibit polarization in the general direction, consistent with the point group. However, in phase VI$_c$, the polarization is confined to the $ab$ plane. As discussed in previous theoretical papers, the emergence of electric polarization induced by noncollinear magnetic ordering can be explained by the inverse DM effect [4,22,25]. When a crystal possesses a mirror plane containing $r_{ij}$ [a vector connecting two spins, $S_i$ and $S_j$, as illustrated in Fig. 16(a)] or a twofold rotation axis perpendicular to $r_{ij}$ exists, as in the orthorhombic manganites with the $Pbnm$ space group, the local electric dipole moment can be described by the formula $p \propto r_{ij} \times [S_i \times S_j] (\equiv p_1)$ [4]. In the absence of these symmetry elements, the additional polarization component expressed by $p \propto S_i \times S_j (\equiv p_2)$ would be expected, as proposed by Kaplan and Mahanti [25]. The ferroaxial mechanism proposed by Johnson $et\ al.$ can also explain polarization parallel to $S_i \times S_j$ in the case of the ferroaxial class [22]. In the case of $\alpha$-NaFeO$_2$ with $R\bar{3}m$, the incommensurate orders at low temperature will always break the threefold rotational symmetry and lower the symmetry to monoclinic $C2/m$. It is convenient to use the extended $k$-vector group to discuss the symmetry-allowed components of the spin-induced polarization [16,42]. Since the $C2/m$ space group possesses neither a mirror plane containing $r_{ij}$ nor a twofold rotation axis perpendicular to $r_{ij}$, the additional term $p_2$ is applicable in addition to $p_1$ for $\alpha$-NaFeO$_2$, just as for other delafossites [16,42]. Figure 16 illustrates the relationship between the noncollinear magnetic structure and the directions of the electric polarization components $p_1$ and $p_2$. Here, we ignore the slight incommensurability of the $a$ and $c$ components in $k_{\text{ICM2}}$.

In the general spiral structure phase VI$_{ab}$, the spin rotation plane is tilted from the $ac$ plane, and therefore the spin helicity, $S_i \times S_j$, has two components: along the $b$ axis (the $y$ axis in Fig. 16), $[S_i \times S_j]_y$, and along the $z$ axis, $[S_i \times S_j]_z$. The $[S_i \times S_j]_y$ component does not generate $p_1$ due to $[S_i \times S_j]_y || y$, while $[S_i \times S_j]_z$ does generate $p_2$. In the same manner, the $[S_i \times S_j]_z$ component results in both $p_1$ along the $x(a)$ axis and $p_2$ along the $z(c^*)$ axis. Therefore, the experimental observation of polarization pointing to a general position can be explained by the extended inverse-DM mechanism including the additional components $p_1$ and $p_2$ [25]. For phases $IV_c$ and $V_c$, an $ab$-cycloid spin structure with $[S_i \times S_j]_z$ produces polarization along the general direction, as shown in Fig. 16(b).

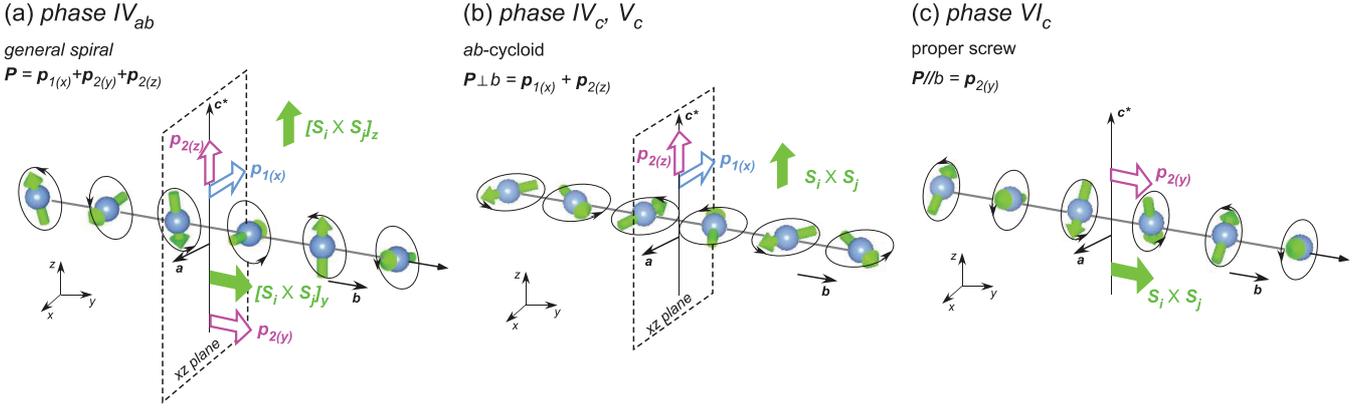

FIG. 16. Schematic illustrations representing the relationships between noncollinear spin modulation along the $b$ axis and the electric polarization directions determined by the extended inverse-DM mechanism [25], $p_1 \propto r_{ij} \times [S_i \times S_j]$ and $p_2 \propto S_i \times S_j$ for each ferroelectric phase: (a) $IV_{ab}$, (b) $IV_c$ and $V_c$, and (c) $VI_c$. The Cartesian coordinates $x$, $y$, and $z$ are defined to be along the monoclinic $a$ and $b$ directions and the axis perpendicular to the $ab$ plane, respectively.

In contrast, the observed electric polarization in phase $VI_c$ is in the $ab$ plane. Taking into account the inverse-DM effect and the proper screw magnetic structure, we can anticipate a $p_2$ component along the $b$ axis. However, based on the small incommensurability of the $a$ and $c$ components of the $k$ vector $[k_{ICM2} = (q_a, q_b, q_c)]$ in the $VI_c$ phase, the symmetry is reduced to triclinic, leading to a $1\,1'$ magnetic point group that allows polarization in the general direction. The incommensurability in the $a$ and $c$ directions generates additional cycloid spin modulations along those directions over significant periods ($\sim 50$ sites and $\sim 100$ sites along the $a$ and $c$ directions, respectively), which creates further electric polarization components along the $z$ and $y$ directions, respectively. However, the absolute values of spin helicity, $|S_i \times S_j|$, generated by modulations along the $a$ direction ($c$ direction) are one order (two orders) of magnitude smaller than that along the $b$ direction. Therefore, although the symmetry allows polarization along the general direction in phase $VI_c$, the out-of-plane polarization component generated by the very long period spin modulations along the $a$ direction can be very small compared to the experimental accuracy in the present polarization measurements.

### B. Exchange interactions

We observed three types of $k$ vectors in the rich phase diagrams of $\alpha$-NaFeO$_2$. These were $k_{CM} = (0.5, 0, 0.5)$ in phase III, $k_{ICM1} = (0, q, \frac{1}{2})$ in phase I, and $k_{ICM2} = (q_a, q_b, q_c)$ in phases II, $IV_{ab}$, $V_{ab}$, $IV_c$, $V_c$, and $VI_c$. Comparing these results with reports regarding the similar delafossite compounds CuFeO$_2$ and AgFeO$_2$ [41,42], we find that the incommensurate $k_{ICM1}$ in $\alpha$-NaFeO$_2$ is common to all the delafossites, in contrast to the commensurate $k = (0, \frac{1}{2}, \frac{1}{2})$ in the case of CuFeO$_2$. The magnetic orderings in CuFeO$_2$ are well explained by the spin Hamiltonian, including "antiferromagnetic" NN exchange interactions, $J_1 < 0$ [42,43]. In contrast, the NN interaction is predicted to be "ferromagnetic," $J_1 > 0$, in $\alpha$-NaFeO$_2$ [29,33].

To roughly estimate the possible magnetic structure of $\alpha$-NaFeO$_2$ as one sublattice and isotropic case, we calculated the Fourier transform of the exchange interactions as

$$E_q = -M^2 J_q = -M^2 \sum_j^{4th} e^{i q \cdot R_j} J_j, \quad (1)$$

where $M$ is the magnitude of the magnetic moment. Here we take the exchange model with exchange interactions up to the fourth NN interactions, including $J_1$ and $J_2$ in the $ab$ plane and interplane $J_{z1}$ and $J_{z2}$, as illustrated in Fig. 17(a). We identified a narrow region in the exchange interaction parameter space in which $E_q$ shows a minimum at $q = (0.5, 0, 0.5)$ when $J_1 > 0$, $J_2/J_1 \simeq -1.0$, $J_{z1}/J_1 \simeq 0$, and $J_{z2}/J_1 \simeq -0.1$. The phase diagram representing the spin state associated with the minimum $E_q$ is shown in Figs. 17(a) and 17(b). The $k_{CM} = (0.5, 0, 0.5)$ lies in the ICM phase in the phase diagram with various $k$ vectors, suggesting that a large number of spin states compete with one another due to spin frustration.

We also investigated the stability of the observed incommensurate $k_{ICM1}$ following the selection of a set of exchange interactions. As shown in the bottom two parts of Fig. 17(b), $E_q$ exhibits a local minimum around $(q_a, 0.24, \frac{1}{2})$ with $q_a \sim 0$, and the $E_q$ value is very close to the global minimum, such that $E_q/|E_q^{min}| = -1$. These results indicate that the spin state with $k_{CM}$ is almost degenerate with the spin state $k_{ICM1}$. When the value of $J_2/J_1$ is changed to $-0.6$, we obtain a local minimum at $(0, 0.24, \frac{1}{2})$, as presented in the inset at the bottom left of Fig. 17(b). Therefore, we can expect that a slight external perturbation, such as the application of a magnetic field, will readily induce the spin state with $k = (0, 0.24, \frac{1}{2})$.

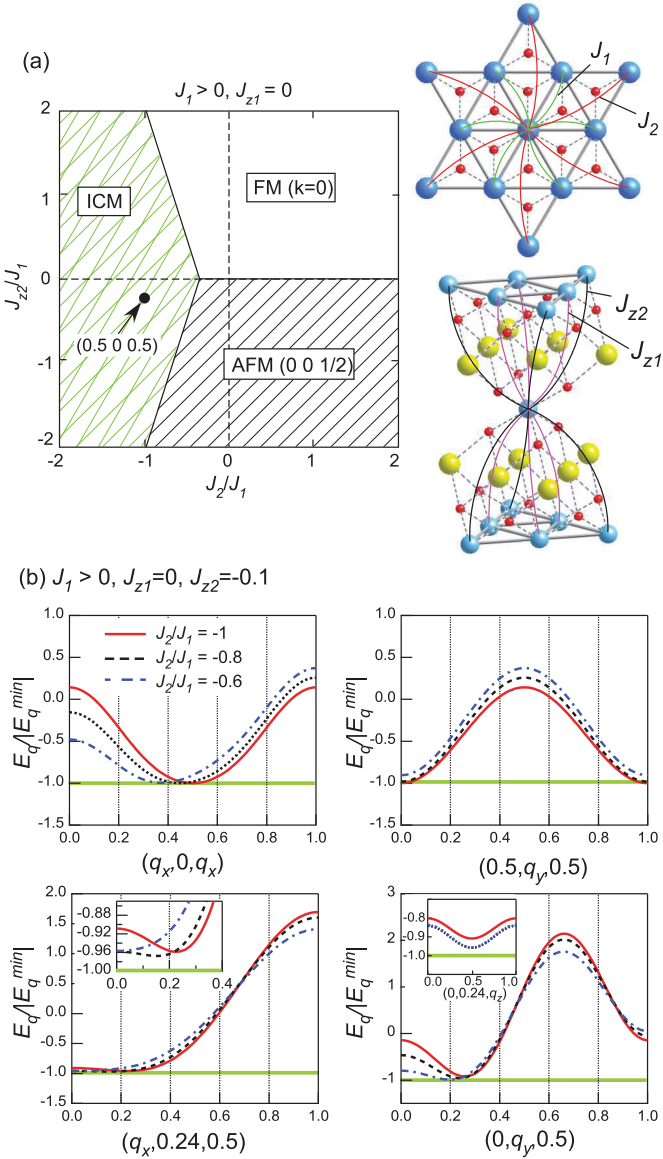

FIG. 17. (a) Magnetic phase diagram of the stable state calculated assuming the ferromagnetic nearest-neighbor exchange constant, $J_1$, and the interplane exchange constant, $J_{z1}$, to be zero. Here, the horizontal and vertical axes are the second-nearest-neighbor interaction, $J_2/J_1$, and the second-nearest interplane interaction, $J_{z2}/J_1$, respectively. The exchange interaction paths are illustrated on the right. (b) Fourier transforms of exchange interactions as functions of typical lines in the reciprocal-lattice space. The inset on the bottom-left figure denotes magnification around (0,0.24,0.5).

This phenomenon is responsible for the rich magnetic phase diagram generated for $\alpha$-NaFeO$_2$. Consequently, although the NN exchange interaction of this material is ferromagnetic (unlike that of other delafossites), the spin state with the same type of incommensurate $k$ vector can be explained using a spin model with exchange interactions up to the fourth NN. However, these calculations do not explain the origin of the incommensurability for the $a$ and $c$ components of $k_{ICM2} = (q_a, q_b, q_c)$. Further theoretical calculations are evidently required to resolve this origin. It should also be noted that the observed magnetization processes along $B_{ab}$ and $B_c$ are completely different from those of the delafossites with antiferromagnetic NN interactions in CuFeO$_2$ [44–46] and AgFeO$_2$ [47], including 1/5 and 1/3 magnetization plateaus. These findings also suggest that significantly different exchange interactions occur in $\alpha$-NaFeO$_2$.

## V. SUMMARY AND CONCLUSIONS

We studied the magnetic and dielectric properties of the multiferroic triangular lattice antiferromagnet $\alpha$-NaFeO$_2$ by macroscopic measurements and neutron diffraction experiments on single crystals grown by hydrothermal synthesis. From the magnetization, specific heat, dielectric permittivity, and pyroelectric current data, we obtained rich magnetoelectric phase diagrams, including five phases in $B_{ab}$ and six phases in $B_c$. No electric polarization was observed in zero magnetic field for phases I, II, and III, while the application of a magnetic field induces polarization along the general direction in phases IV$_{ab}$, V$_{ab}$, IV$_c$, and V$_c$, and in the $ab$ plane for phase VI$_c$. Neutron diffraction experiments under $B_{ab}$ and $B_c$ are in agreement with the phase transitions, based on changes in the $k$ vector and the intensities of magnetic reflections. Three types of $k$ vectors were observed: $k_{ICM1} = (0, q, \frac{1}{2}; q \simeq 0.24)$ in phase I; $k_{ICM2} = (q_a, q_b, q_c; q_a \simeq 0.02, q_b \simeq 0.24, q_c \simeq 0.49)$ in phases II, IV$_{ab}$, V$_{ab}$, IV$_c$, V$_c$, and VI$_c$; and $k_{CM} = (0.5, 0, 0.5)$ in phase III. The three components $q_a$, $q_b$, and $q_c$ in $k_{ICM2}$ depend greatly on the temperature and magnetic fields in the case of each of the phases, while $k_{ICM1}$ and $k_{CM}$ are field- and temperature-independent. Based on a magnetic structure analysis with symmetry considerations, we determined the magnetic structures and $(3+1)$ superspace groups for each magnetic phase, as summarized in Table I. In the ferroelectric phases, the general spiral in IV$_{ab}$ and the $ab$ cycloid in IV$_c$ and V$_c$, with the magnetic point group $11'$ are consistent with the observed polarization in the general direction.

The relationship between the observed polarization and the magnetic structures can also be explained by the extended inverse-DM mechanism [25] with two orthogonal polarization components, $p_1 \propto r_{ij} \times [S_i \times S_j]$ and $p_2 \propto S_i \times S_j$, in phases IV$_{ab}$, IV$_c$, and V$_c$ in $\alpha$-NaFeO$_2$. In the case of phase VI$_c$, the electric polarization appears to be confined to the $ab$ plane, even though the magnetic point group of $11'$ allows polarization in the general direction. The period of spin modulations along the $a$ and $c$ directions is very long, leading to very small additional polarization components along the $c^*$ direction. The above results assist in explaining the observation that out-of-plane polarization could not be observed during measurements of phase VI$_c$. Finally, this work allows a discussion of the expected exchange model for $\alpha$-NaFeO$_2$ by comparing it with the model for similar delafossite triangular lattice systems with antiferromagnetic NN interactions, such as $A$FeO$_2$ ($A$ = Cu and Ag). In our calculations with exchange interactions up to the fourth NN, ferromagnetic NN $J_1 > 0$, antiferromagnetic $J_2/J_1 \simeq -1.0$, interplane $J_{z1}/J_1 \simeq 0$, and $J_{z2}/J_1 \simeq -0.1$, we found that the most stable state, $k = (0.5, 0, 0.5)$, in this model was almost degenerate with the incommensurate state $k = (0, q, \frac{1}{2})$. In spite of the unique ferromagnetic NN exchange interaction in the present case (which is unlike those of other delafossites), the spin state with the same type of incommensurate $\boldsymbol{k}$ vector can be explained based on an exchange model up to the fourth NN interaction. Considering the set of exchange couplings leading to degenerate spin states, we infer that the rich phase diagram of $\alpha$-NaFeO$_2$ can also be attributed to very strong competition among the antiferromagnetic second NN interactions as well as weakly coupled interplane antiferromagnetic interactions.

TABLE I. Summary of the $\boldsymbol{k}$ vectors, magnetic structures, superspace groups (magnetic space group only for phase III), point groups, and electric polarization directions observed for $\alpha$-NaFeO$_2$. We did not determine the magnetic structure for phase V$_{ab}$.

| Phase | $\boldsymbol{k}$ vector | Magnetic structure model | (Super)space group | Point group | $P$ direction |
|---|---|---|---|---|---|
| I | $(0,q,\frac{1}{2})$ | SDW ($S_{\|ac}$) | $C2/m1'(0,\beta,\frac{1}{2})s0s$ | $2/m1'$ | |
| II | $(q_a,q_b,q_c)$ | SDW ($S_{\|general}$) | $P\bar{1}1'(\alpha,\beta,\gamma)0s$ | $\bar{1}1'$ | $P=0$ |
| III | $(0.5,0,0.5)$ | Collinear ($S_{\|b}$) | $P_a2_1/m$ | $2/m$ | |
| IV$_{ab}$ | | General spiral ($S_{\|general}$) | | | |
| V$_{ab}$ | | | | | $P_{\|general}$ |
| IV$_c$ | $(q_a,q_b,q_c)$ | $ab$ cycloid ($S_{\|ab}$) | $P11'(\alpha,\beta,\gamma)0s$ | $11'$ | |
| V$_c$ | | | | | |
| VI$_c$ | | Proper screw ($S_{\|ac}$) | | | $P_{\|ab}$ |

## ACKNOWLEDGMENTS

The images shown in Figs. 10, 12, 15, 16, and 17 were generated using the VESTA [48] software program developed by K. Momma. This work was supported by a JSPS KAKENHI Grant (No. 15H05433) and by the TUMOCS project, which has received funding from the European Union Horizon 2020 Research and Innovation Program under the Marie Sklodowska-Curie Grant Agreement (No. 645660).